\documentclass[%
 reprint,
 amsmath,amssymb,
 aps,
]{revtex4-2}

\usepackage{graphicx}
\usepackage{dcolumn}
\usepackage{bm}
\usepackage{xcolor}
\usepackage[normalem]{ulem}

\begin{document}

\title{Energy Conversion and Entropy Production in Biased Random Walk Processes - from Discrete Modeling to the Continuous Limit}

\author{Henning Kirchberg}
\email{khenning@sas.upenn.edu}
\author{Abraham Nitzan}
\affiliation{
 University of Pennsylvania, Department of Chemistry, Philadelphia, PA, 19104
}

\date{\today}

\begin{abstract}
We consider discrete and continuous representations of a thermodynamic process in which a random walker (e.g. a molecular motor on a molecular track) uses a periodically pumped energy (work) to pass $N$ sites and move energetically downhill while dissipating heat. Interestingly, we find that, starting from a discrete model, the limit in which the motion becomes continuous in space and time ($N\to \infty$) is not unique and depends on what physical observables are assumed to be unchanged in the process. In particular, one may (as usually done) choose to keep the speed and diffusion coefficient fixed during this limiting process, in which case the entropy production is affected. In addition, we study also processes in which the entropy production is kept constant as $N\to \infty$ at the cost of modified speed or diffusion coefficient. Furthermore, we also combine this dynamics with work against an opposing force, which makes it possible to study the effect of discretization of the process on the thermodynamic efficiency of transferring power input to power output. Interestingly, we find that the efficiency is increased in the limit of $N\to\infty$. Finally, we investigate the same process when transitions between sites can only happen at finite time intervals and study the impact of this time discretization on the thermodynamic variables as the continuous limit is approached.
\end{abstract}

\maketitle

\section{Introduction}
\label{intro}

Non-equilibrium thermodynamics deals with general laws of a (driven) system transferring energy from one or more heat bath(s) to useful work. The second law, however, restricts this transformation as only part the input energy may be "accessible" as the entropy production, related to heat production, must not decrease \cite{cla1854,bol1877,sei2012}. The "system" is usually described by many degrees of freedom, e.g., $10^{23}$ gas particles with individual fluctuating trajectories, where the exchanged heat and the extracted work are determined from statistical averages over all these particles trajectories \cite{SekimotoBook}. For "small" systems, such as (bio)polymers, colloid particle, enzymes or molecular motors, the dynamics is described by only few degrees of freedom and the fluctuation of individual "state" trajectories becomes more prominent \cite{sei2011,sei2012,ein2016}. Notably, it has been shown, e.g., see Refs. \cite{bar2015,pie2016,dec2018}, that the resulting (relative) fluctuations of a statistical averaged observable, like the number of steps $R$ of a molecular motor mimicked by a random walk along a track of molecules, can be related to the entropy production. A relative uncertainty or fluctuation $\Delta R^2 = 2D/v^2$ of the motor steps along the molecular track with diffusion coefficient $D$ and velocity $v$ requires at least a entropy production rate $\dot{\sigma}$ of $2k_B/\Delta R^2$. This leads to the inequality, known as the thermodynamic uncertainty relation (TUR) \cite{bar2015,pie2016}, 
\begin{align}
\label{TUR}
\dot{\sigma}\frac{2 D}{v^2}\geq 2k_B.
\end{align} 

Different approaches are used for describing small system dynamics including fluctuations and the resulting (stochastic) thermodynamic properties. Two of these approaches
have been prominently elaborated in recent years \cite{sei2012}: 

(i) \textit{Dynamics on a discrete set of states}: A system is described by its \textit{micro}states whose dynamics is captured by a \textit{master equation}. The rates to interchange between microstates is governed by local detailed balance relation determined by a thermal bath \cite{sei2012}. 

(ii) \textit{Dynamics on continuous trajectories}: Consideration of individual continuous trajectories of a (driven) colloidal particle whose velocity is described by a Langevin equation where the probability distributions of a particle at position $x$ with velocity $v$ and diffusion $D$ are determined by the \textit{Fokker-Planck equation}.

Starting from discrete set of states, we can approximate the dynamics by a continuous description, such as the Fokker-Planck equation by standard procedures like the Kramers-Moyal expansion, when (infinite) many states may be visited during the time of observation, e.g., a chemical reaction network of multiple reaction steps \cite{VanKampenBook,kra1940,moy1949,RiskenBook,NitzanBook}. 
Similarly, starting from a continuous description for the dynamics, a discrete representation is obtained by standard mathematical steps of replacing derivatives by finite difference quotients. While such mathematical transformations are expected to lead to equivalent descriptions of the underlying physics, we show below that not all physical observables can be kept invariant under such transformations. In particular, we show that the discrete to continuous limiting process is not unique and depends on which observables are chosen to be invariant under this process. Recent studies have already shown that the entropy production might differ under different coarse-graining schemes since under coarse-graining some "information" is lost while also the mathematical derivation of the differential entropy starting from the discrete Shannon entropy has revealed some discrepancy \cite{footnote5,jan1963,esp2012,bus2019}. Here we suggest a different point of view by showing that it is possible to impose an invariant entropy production (or a given heat exchanged with the thermal environment) when proceeding from the discrete to the continuous description of the system dynamics. Under such restriction some other system observable cannot be kept invariant. Alternatively, one may even ask if this observation may be translated into realistic system processes. Put differently, can we, by adding intermediate states between an initial and a target state, optimize the process of transferring input energy into useful work.

The aim of the present manuscript is to investigate the transition from a discrete to a continuous state-space system for an exactly solvable master equation by keeping distinct system observables constant while studying the impact on other observables. Explicitly, we consider a simple model: A Brownian particle moving on a downhill slope with an energy pumping step taking place at constant length intervals $L$ that restores the energy of the particle into its initial value which represents a cycle. In the discrete representation, the particle is a random walker moving among $N$ equally spaced sites per fixed length $L$. $N\to \infty$ that represents the continuous limit for the cyclic dynamics. More details about the model are provided in Sec.\ \eqref{unicylcic}. Other processes with period boundary conditions can be mapped into this form. Realization of such a process can be molecular motors which transform free energy liberated in a chemical reaction by a succession of steps on a track into mechanical work (motion) \cite{sei2011,pie2016}. Driven rotational Brownian particles through periodic potential wells for experimentally testing thermodynamics laws represent another example of many others \cite{toy2010}. Given the specific chosen conditions, like constant speed $v$ and the entropy production rate of the cycle, we can study the diffusion coefficient $D$ in dependence of the number of sites $N$ per cycle and the continuous limit when $N\to \infty$. Interestingly, under an additional action of an opposing force during the process under study, we can also determine the impact of the number of sites per cycle on thermodynamic performance by transferring input energy into useful work. Furthermore, when keeping the energy drop per cycle and the speed constant, we interestingly find that in the limiting case $N \to \infty$ the diffusion coefficient (equivalent to the variance in the site distribution) approaches the thermal Einstein relation when $N \to \infty$ usually expected only for small velocities in the linear response limit of small driving. 

The paper is structured as follows. In the first Section \ref{unicylcic}, we describe the cyclic process as biased random between (energy) sites governed by a master equation restricted by periodic boundary conditions. We further discuss the implication for the entropy and heat production. In Sec.\ \ref{SecIII}, we study the entropy production rate and heat transfer rate into the environment of the cyclic process given the velocity $v$ and diffusion constant $D$ for different number of sites $N$ in the dynamical (relaxation from initial site) and in steady state. We then determine
the performance in transferring input power to useful output power for the cycle process under an opposing force by varying $N$ and, so, by going from the discrete to the continuous ($N\to \infty$) state space. We study in Sec.\ \ref{IV} the transition to $N \to \infty$ given constant entropy production for the process, and, under either constant velocity $v$ or diffusion coefficient $D$. The impact on the respective system variable when increasing $N$ is discussed. Section \ref{randomness} is devoted to examining the randomness parameter of the cycle, which is dependent on the number of sites $N$, while maintaining various physical observables as constants. Through this randomness factor, we can also determine the variance in cycle completion time and study its dependence on $N$. In Sec.\ \ref{V}, we investigate the process by the same biased random walk, but where the system evolves at finite time intervals, and study the impact of this time discretization on the entropy production and diffusion coefficient. Sec.\ \ref{conc} concludes this work.
 
\section{Cyclic process}
\label{unicylcic}
We consider a cyclic process, where the $n=1,\dots,N$ sites are aligned on a circle such that each site has two neighboring sites with the periodic boundary condition $N+n=n$ with $N\geq3$. The process dynamics is captured by a biased random walk with forward and backward transfer rates $\alpha$ and $\beta$, which are equal at each site, such that the master equation for the probability distribution of the $n$-site of the cycle reads
\begin{align}
\label{master1}
\dot{P}(n,t)=\alpha P(n-1,t) + \beta P(n+1,t)-(\alpha+\beta)P(n,t).
\end{align}
Starting form a well defined site at $t=0$ one finds in the limit of long times $t\to \infty$ the (nonequilibrium) steady state distribution for the $n$ site to be $P(n,t\to \infty)=1/N$ (see Appendix \ref{N-site cycle}). When writing the master equation \eqref{master1} in the form $\dot{\mathbf{P}}=M\mathbf{P}$ we see that the matrix $M$ is irreducible with the dominant eigenvalue $\lambda=0$ while all other eigenvalues have a strictly negative real part according to the Perron-Frobenius theorem signaling the existence of a stable steady state vector \cite{per1907,fro1912}. While our cyclic process can take place in the state space of any given system, it is convenient to consider the equivalent random walk with forward and backward rates $\alpha$ and $\beta$ on a cycle of constant circumference $L=2\pi R$. In this equivalent random walk problem the velocity $v$ and diffusion constant $D$ at steady state are defined by \cite{NitzanBook} 
\begin{align}
\label{eq9}
v\equiv \frac{\langle n \rangle \Delta x}{t}=(\alpha-\beta)\Delta x
\end{align}
and
\begin{align}
\label{eq9b}
2D\equiv \frac{(\langle n^2 \rangle- \langle n \rangle^2)\Delta x^2}{t}=(\alpha+\beta)\Delta x^2,
\end{align}
where $\Delta x = 2\pi R /N$ is the equidistant step size with $R$ and $N$ being the radius of the cycle and the number of total sites $N$ of the cycle respectively. Evidently, the time for completing a full cycle is $\tau=N(\alpha-\beta)^{-1}$.

The transition between the neighboring sites, $n\to n\pm 1$, are considered as autonomous Markov jump processes where each site has its energy $E(n)$. Thermodynamic consistency is introduced by the local detailed balance condition
\begin{align}
\label{eqLocal}
\frac{\alpha}{\beta}=e^{\frac{1}{k_BT}\big(E(n)-E(n+1)\big)},
\end{align}
where, for simplicity, we assume isothermal conditions with $T$ on all sites, and where $\Delta E= E(n)-E(n+1)>0$ is taken the same for all nearest neighbor sites. Note that we count the heat exchange with the bath as $Q\equiv -\Delta E$, i.e., if $Q<0$ the amount of heat is taken from the system to the bath while the environment provides heat to the system, if $Q>0$. This implies that during a cycle, the amount of heat $\Delta Q\equiv -N \Delta E=k_BT N\ln\big(\frac{\alpha}{\beta}\big)$ is dissipated, so $E(N+1)\equiv E(1)-N\Delta E$ may be compared to a downhill process of energy loss $N\Delta E$. To remain consistent with the periodic boundary conditions we further assume that between sites $N$ and $N+1=1$ an upward energy jump occurs, in which the same amount of work, $W=N \Delta E$, is returned to the system. 

Next we calculate the time-dependent change of the Boltzmann-Gibbs entropy, $\dot{S}(t)=-k_B \sum_n \dot{P}(n,t) \ln P(n,t)$, \cite{bol1877,GibbsBook} for the system using the master equation \eqref{master1}. We obtain
\begin{align}
\dot{S}(t)&= \label{eq1a}-k_B\sum_{m=1}^N\sum_{n=1}^N \big[\alpha P(n-1,t) + \beta P(n+1,t)\\ \notag &-(\alpha+\beta)P(n,t)\big] \ln{\frac{P(n,t)}{P(m,t)}}
\\ &=k_B\sum_{m=1}^N\sum_{n=m-1}^{N-1} \big[\alpha P(n,t) - \beta P(m,t)\big] \ln{\frac{P(n,t)}{P(m,t)}}
\\ \label{eq2a}  &= k_B\sum_{m,n=m-1}  \big[\alpha P(n,t) - \beta P(m,t)\big] \ln{\frac{P(n,t)\alpha}{P(m,t)\beta}}
\\ \notag &-k_B\sum_{m,n=m-1}  \big[\alpha P(n,t) - \beta P(m,t)\big] \ln{\frac{\alpha}{\beta}}.
\end{align}
Note that in Eq.\ \eqref{eq1a} we could replace $\ln P(n,t)$ by $\ln[P(n,t)/P(m,t)]$ because the sum that multiply $\ln P(m,t)$ vanishes. 
The result \eqref{eq2a} can be recast in the form $\dot{S}(t)=\dot{S}_e+\dot{\sigma}(t)$, \textcolor{blue}{\cite{sch1976,esp2010}} where, using $\sum_nP(n,t)=1$, 
\begin{align}
\label{eq2bb}\dot{S}_e&=-k_B\sum_{m,n=m-1}  \big[\alpha P(n,t) - \beta P(m,t)\big] \ln{\frac{\alpha}{\beta}} \\ \notag &=-k_B \big[\alpha - \beta \big] \ln{\frac{\alpha}{\beta}},
\end{align}
and
\begin{align}
\label{eq2b}\dot{\sigma}(t)&=k_B\bigg[ \big[\alpha- \beta \big] \ln{\frac{\alpha}{\beta}} \\ \notag &+\sum_{m,n=m-1}  \big[\alpha P(n,t) - \beta P(m,t)\big] \ln{\frac{P(n,t)}{P(m,t)}}\bigg].
\end{align}

Next we show that first term $\dot{S}_e$ (Eq.\ \eqref{eq2bb}) is the entropy flow into the environment while the second term is entropy production rate $\dot{\sigma}(t)$ (Eq.\ \eqref{eq2b}).

Consider first Eq.\ \eqref{eq2bb}. Because the rates $\alpha$ and $\beta$ were assumed not depend on the site identity, $\dot{S}_e$ is time-independent. $\dot{S}_e$ can be written as
\begin{align}
\label{eq4}
\dot{S}_e= J \frac{Q}{T},
\end{align}
where $J\equiv \alpha-\beta$ is the cumulative flux: The sum over nearest neighbor site-pair fluxes (see first expression in Eq.\ \eqref{eq2bb}). $Q$ is the heat exchanged during a single nearest neighbor transfer event with the environment according to the local detailed balance relation in Eq.\ \eqref{eqLocal}. 
The product $J Q$ in Eq.\ \eqref{eq4} is the heat flux into the thermal environment of temperature $T$ per cycle. For the given model this heat exchange is time-independent. 

According to the Clausius principle the change in system entropy is bounded by the (negative) heat amount exchanged with the environment $\dot{S}\geq  \frac{JQ}{T}$ where equality is reached for reversible processes\cite{cla1854}. Motivated by this inequality, one defines the total entropy production rate by $\dot{\sigma}(t)=\dot{S}-\frac{JQ}{T}\geq 0$. 
Indeed, the second term $\dot{\sigma}(t)$ (Eq.\ \eqref{eq2b}) meets the two important properties: (i) It is nonegative because the firs term in Eq.\ \eqref{eq2a} can be recast into $(x-y)\ln(x/y)\geq 0$ and (ii) it vanishes for thermal equilibrium, when microscopic reversibility or detailed balance condition, $\alpha P(m,t)=\beta P(n,t)$, is obeyed and no entropy is produced. At (nonequilibrium) steady state Eq.\ \eqref{eq1a} is zero and the entropy production equals the negative entropy flow into the environment $\dot{\sigma}=-\dot{S}_e$, Ref. \cite{tom2005,esp2010}.

In the following, we investigate the entropy production rate $\dot{\sigma}(t)$ and the physical measurable entropy or heat flow $\dot{S}_e$ into the environment for prototype $N$-site cyclic process given different conditions. In particular we are interested in the limit $N\to \infty$ to investigate the transition from the discrete to the continuous state-space. 

\section{N-site cyclic process under constant velocity $v$ and diffusion constant $D$}
\label{SecIII}

We first consider the dependence of the number of sites $N$ per cycle on the system that is done under the conditions that (a) the speed $v$ (Eq.\ \eqref{eq9}) and (b) the diffusion constant $D$ (Eq.\ \eqref{eq9b}) are kept constant. The first condition requires that the time, $\tau =N(\alpha-\beta)^{-1}$, for a full completion of the cycle remains constant.
Together with the second condition, the forward and backward rate must depend on $N$. We get
\begin{align}
\label{eq12}
\alpha&=\frac{DN^2}{4\pi^2R^2}+\frac{vN}{4\pi R},\\
\label{eq13}
\beta&=\frac{DN^2}{4\pi^2R^2}-\frac{vN}{4\pi R}.
\end{align}
To ensure the positivity of the rate \eqref{eq13}, $\beta \geq 0$, it immediately follows that $v\leq N D /R\pi$. Note that under the conditions restricted by Eqs.\ \eqref{eq12} and \eqref{eq13}, the detailed balance relation $\Delta E =k_B T \ln \big( \frac{\alpha}{\beta} \big)$ (Eq.\ \eqref{eqLocal}) will be a nonlinear function of $N$. 

\begin{figure}[h!!!!!]
\centering
\includegraphics[width=\linewidth]{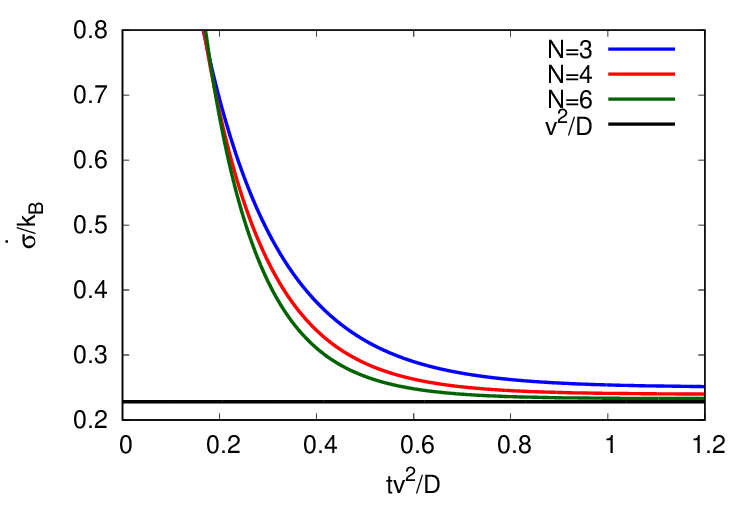}
\caption{\label{fig1} Entropy production rate $\dot{\sigma}(t)$ against time plotted for different $N$-site cycle by keeping the velocity $v$ and diffusion coefficient $D$ constant. The velocity $v$ (here chosen to be $v=\frac{3D}{2\pi R}$) needs to be within the bounds implied by $\beta\geq 0$ of Eq.\ \eqref{eq13} for $N\geq 3$. The steady state value $v^2/D$ (black curve) is reached for $N\to \infty$.}
\end{figure}

Consider the entropy production rate $\dot{\sigma}(t)$ (Eq.\ \eqref{eq2b}) given a well define initial cycle site $n=1$ at $t=0$, so $P(n=1,t=0)=1$. The entropy production rate (the entropy produced per unit time) decreases over time before reaching a ($N$-dependent) steady state value, see Fig.\ \ref{fig1}. 
This observation can be understood as follows. Initially, the system starts at a given site and evolves in its cyclic dynamics according to the master equation \eqref{master1}. At later time $t$ the system will be found at a given site $n$ with probability $P(n,t)$ (see Appendices \ref{three}-\ref{N-site cycle}). The loss of the initially "knowledge" about the exact system site $n$ increases the entropy of the system. As the probability distribution relaxes to its steady state distribution $P_{n,SS}\equiv P(n,t\to \infty)=1/N$ (see Appendix \ref{N-site cycle}) the entropy production rate, $\dot{\sigma}$, decreases to its steady state value 
\begin{align}
\label{heatsteadystate}
\dot{\sigma}_{SS}&=k_B (\alpha - \beta) \ln{\bigg [ \frac{\alpha}{\beta}\bigg ]}
=\frac{k_B N}{2 \pi R}  v  \ln{\bigg [ \frac{\frac{2D N}{2 \pi R} + v }{\frac{2 D N}{2 \pi R} -  v }\bigg ]}.
\end{align}

Note that entropy is produced at constant rate when running the cyclic process under steady state conditions. 
Interestingly, as seen in Fig.\ \ref{fig1}, the more sites $N$ are included in the cycle of finite length the faster the entropy production rate decreases to its steady state value. At steady state the entropy production equals the negative (in this model time-independent) entropy flow into the environment $\dot{\sigma}_{SS}=-\dot{S}_e$.

\begin{figure}[h!!!!!]
\centering
\includegraphics[width=\linewidth]{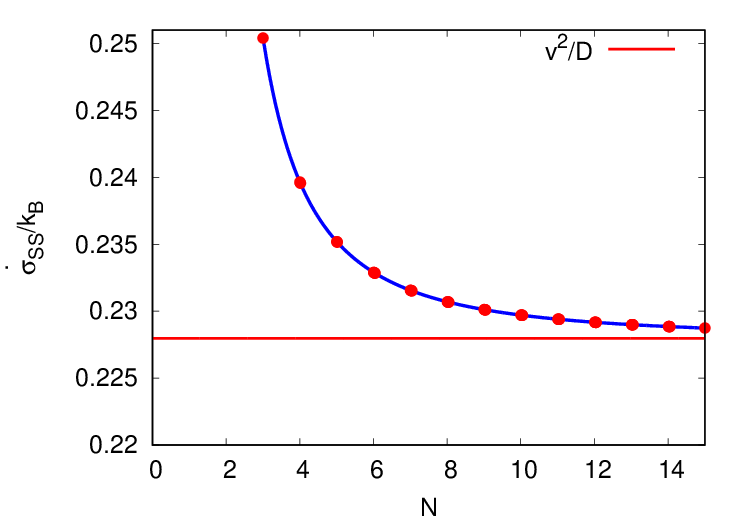}
\caption{\label{fig2} Entropy production rate at steady state $\dot{\sigma}(t\to \infty)\equiv \dot{\sigma}_{SS}[=-\dot{S}_e]$ against the number of sites $N$ per cycle by keeping the velocity $v$ and diffusion coefficient $D$ constant. The velocity $v$ (here chosen to be $v=\frac{3D}{2\pi R}$) needs to be within the bounds implied by $\beta\geq 0$ of Eq.\ \eqref{eq13} for $N\geq 3$. The red line is the entropy production rate in the limit $N\to \infty$ and takes the value $\dot{\sigma}_{SS}(N\to \infty)=v^2/D$.}
\end{figure}

Consider the $N$-dependence of the the steady state entropy production rate $\dot{\sigma}_{SS}$ for an $N$-site cycle, Eq.\ \eqref{heatsteadystate}. The dependence is shown in Fig.\ \ref{fig2}. For $N\to \infty$ the steady state entropy production rate \eqref{heatsteadystate} is reduced to $k_B v^2/D$. This minimum entropy production rate in this limit can be understood as the forward and backward rates (Eqs.\ \eqref{eq12} and \eqref{eq13}) will become more alike. Therefore, consecutive transitions between sites $n$ and $n+1$ become more time-symmetric and less entropy per step, $\Delta E / T=\ln{[\alpha(N)/\beta(N)]}/ T$, is produced.

Interestingly, for $N\to \infty$, the TUR relation, the minimal required entropy production rate $\dot{\sigma}$ for a given relative fluctuation $2D/v^2$ (as introduced in Eq.\ \eqref{TUR} for a molecular motor moving along a molecular track) states equality, $\dot{\sigma}2D/v^2=2k_B$. 
In this limit, the dynamics of the cyclic process is comparable to a continuous Brownian diffusive motion of a particle of constant speed $v$ and diffusion $D$ described by a Fokker-Planck equation \cite{coc2020}.

\begin{figure}[h!!!!!]
\centering
\includegraphics[width=\linewidth]{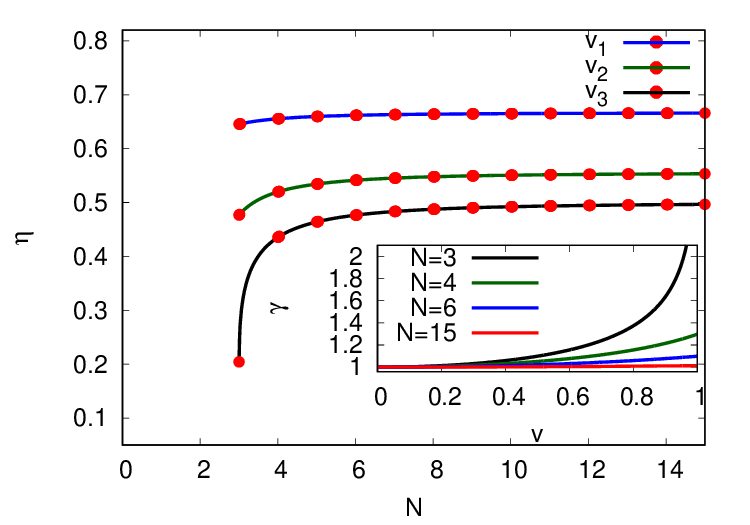}
\caption{\label{fig2b} Efficiency at steady state $\eta$ plotted against the number of sites $N$ per cycle by keeping the velocity $v$ and diffusion constant $D$ constant. We depict $\eta$ for $v_1=0.5v$, $v_2=0.8v$ and $v_3=v$, where we have chosen $v=\frac{3D}{\pi R}$ to be within the bounds implied by $\beta \geq 0$ of Eq.\ \eqref{eq13} for $N\geq 3$. We have chosen the force $f\equiv v T \dot{\sigma}_{SS}(N\to \infty)=3k_BT/\pi R$. Inset: Damping constant $\gamma\equiv \gamma(v)D/(k_BT)=TD\dot{\sigma}_{SS}/(v^2k_BT)$ in dependence of the velocity for different $N$.}
\end{figure}

Assume that a forward (downhill) step on the cycle takes place against a constant applied force $f$. Then, the local detail balance relation \eqref{eqLocal} must be redefined by $\Delta \tilde{E}\equiv E(n)-E(n+1)-f \Delta x=\Delta E -f \Delta x>0$, as part of the energy per step $\Delta E$ is transferred to work $f\Delta x$, where $\Delta x$ is the step size. We defined the system's heat exchange per step with the bath as $Q\equiv -\Delta \tilde{E}$, while $Q<0$ is the amount of heat taken from the system to the bath. At steady state the heat flow into the bath is $-\dot{Q}=T\dot{\sigma}_{SS}$, see discussion in Sec.\ \ref{unicylcic}. The power output per cycle when running against the force at steady state is 
\begin{align}
\label{power}
P\equiv \sum_{m,n=m-1}  & \big[\alpha P(n) - \beta P(m)\big] f \Delta x=(\alpha-\beta)\Delta x f\\ \notag &=v f.
\end{align}
In Eq.\ \eqref{power} $\alpha P(n) - \beta P(n+1)$ is the probability flux between neighboring sites and where $\sum_n P(n)=1$. 
By the first law of thermodynamics the total supplied power must be $P-\dot{Q}$ such that we can define the thermodynamic efficiency as \cite{pie2016}
\begin{align}
\label{efficiency}
\eta(N) &\equiv\frac{P}{P-\dot{Q}} = \frac{fv}{fv+T\dot{\sigma}_{SS}(N)},
\end{align}
where $\dot{\sigma}_{SS}(N)$, Eq.\ \eqref{heatsteadystate}, depends on the total number of sites $N$ per cycle. The cycle can be compared to a process of going down a slope against a constant force and with friction. The friction force is usually taken as $F_{fr}=\gamma v$ such that the related heat dissipated in the cycle per unit time is $F_{fr}v=\gamma v^2$. By identifying $\gamma v^2\equiv T\dot{\sigma}_{SS}(v,N)$ we can calculate the friction coefficient $\gamma(v)=T\dot{\sigma}_{SS}(v)/v^2$ for the present cycle process using Eq.\ \eqref{heatsteadystate}. 
Note that, as expected, the friction coefficient $\gamma$ goes to its linear response value for a small velocity, $\gamma(v \to 0) \to k_BT/D$, independent of $N$. 
Surprisingly, as depicted in the inset of Fig.\ \ref{fig2b}, $\gamma$ takes the same value also for finite $v$ in the limit of $N \to \infty$. 
In this limit the system assumes some 
features similar to equilibrium even though the flux is finite. Not only in the limit $v \to 0$, but also in the limiting process $N \to \infty$ can be compared to thermodynamic cycles where the system changes adiabatically slowly to be always in thermal equilibrium throughout the process.  
 
Equivalently, in the limit $N\to\infty$ we can write the thermodynamic efficiency \eqref{efficiency} as \cite{footnote1}
\begin{align}
\label{efficiency2}
\eta =\frac{1}{1+k_BTv/Df}.
\end{align}

Fig.\ \ref{fig2b} portrays the efficiencies for different $N$. As expected, the efficiency increases with $N$ as less heat $-\dot{Q}=T\dot{\sigma}_{SS}$ will be produced per cycle given constant power output $P$. Interestingly, the slower the velocity $v$ is chosen the more one can reach maximal efficiency. The efficiency is bound from above, $\eta\leq 1$, while equality is reached for $v=0$ or $f\to \infty$, $D\to \infty$, see Eq.\ \eqref{efficiency2}. The last conditions, however, do not produce useful output power, as the cycle will stop.

\section{Cyclic process with constant velocity $v$ or diffusion constant $D$ and constant energy drop per cycle}
\label{IV}

As stated in the introduction, the process of going to the continuous description for the state-space dynamics ($N\to \infty$) is not unique. We therefore use the same methodology as above to describe the unicyclic process as 1-D random walk but where we require now the total entropy produced per cycle is constant (given by the constant energy drop per cycle) and either (A) the velocity $v$ or (B) the diffusion coefficient $D$ are constant.
We assume that all energy invested into the system is dissipated as heat to the environment, so $W=\sigma T$. The steady state entropy production $\sigma$ is
\begin{align}
\sigma= k_B N \ln{\bigg(\frac{\alpha}{\beta}\bigg)},
\end{align}
where $N$ is the total number of sites.
At steady state the entropy production equals the heat going into the environment, see Sec. \ref{unicylcic}. 

(A) Given the constant velocity $v$, Eq.\ \eqref{eq9}, the forward and backward rates are
\begin{align}
\label{a1}
\alpha &= \frac{vN}{2\pi R\big(1-e^{\frac{-\sigma}{Nk_B}}\big)}, \\
\label{a2}
\beta &= \frac{vN}{2\pi R\big(e^{\frac{\sigma}{Nk_B}}-1\big)},
\end{align}
where $R$ is the radius of the cycle (see Sec.\ \ref{unicylcic}).

(B) Given the constant diffusion coefficient $D$, Eq.\ \eqref{eq9b}, the forward and backward rates are
\begin{align}
\label{b1}
\alpha &= \frac{2DN^2}{(2\pi R)^2\big(1+e^{\frac{-\sigma}{Nk_B}}\big)}, \\
\label{b2}
\beta &= \frac{2DN^2}{(2\pi R)^2\big(e^{\frac{\sigma}{Nk_B}}+1\big)}.
\end{align}

\begin{figure}[h!!!!!]
\centering
\includegraphics[width=\linewidth]{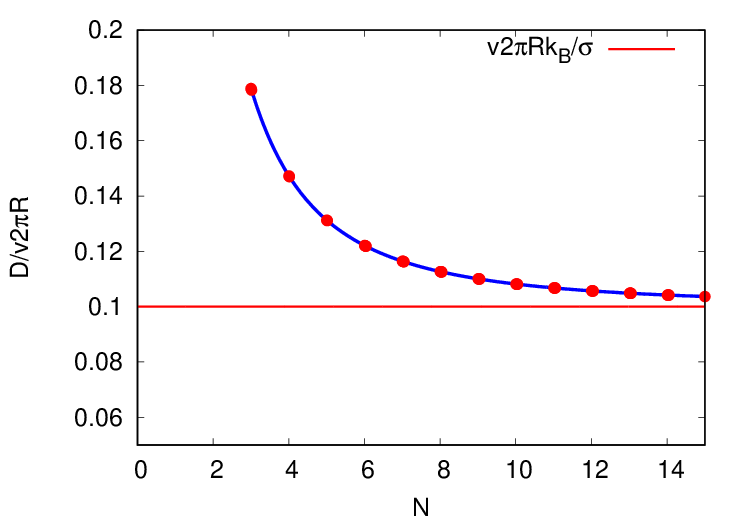}
\caption{\label{fig3} Diffusion constant $D$ plotted against the number $N$ of sites per cycle by keeping the entropy $\sigma$ per cycle and velocity $v$ constant for $N\geq 3$. We choose the produced entropy per cycle to be $\sigma/k_B=10$. The red line is the value $D(N\to \infty)\to v 2\pi R k_B/\sigma$ in the limit of $N\to \infty$.}
\end{figure}

Consider first case (A). Eq.\ \eqref{eq9b} with Eqs.\ \eqref{a1} and \eqref{a2} lead to
\begin{align}
\label{Diffusion}
D&=\frac{1}{2}(\alpha+\beta)\bigg(\frac{2\pi R}{N}\bigg)^2=\frac{v\pi R}{N} \coth{\bigg[ \frac{\sigma}{2Nk_B} \bigg]},
\end{align}
which is shown in Fig.\ \ref{fig3}. The diffusion coefficient decreases with increasing $N$. As the diffusion coefficient is the variance of the site distribution on our equivalent cycle, see Eq.\ \eqref{eq9b}, the related fluctuations in localization of a site is reduced during a cycle with increase of $N$. Assuming that the energy falls uniformly along the cycle, so that $\Delta E=T\sigma/N$, we find that in the regime of linear response $v\to 0$ and in the limit $N\to \infty$ the diffusion coefficient is captured by the (Einstein) relation $D=v/Fk_BT=\mu k_BT$ with the mobility $\mu=v/F$ and the related force $F=T\sigma /2\pi R$ in analogue to the friction coefficient (as discussed in Sec.\ \ref{SecIII})  \cite{ein1905,smo1906}.

\begin{figure}[h!!!!!]
\centering
\includegraphics[width=\linewidth]{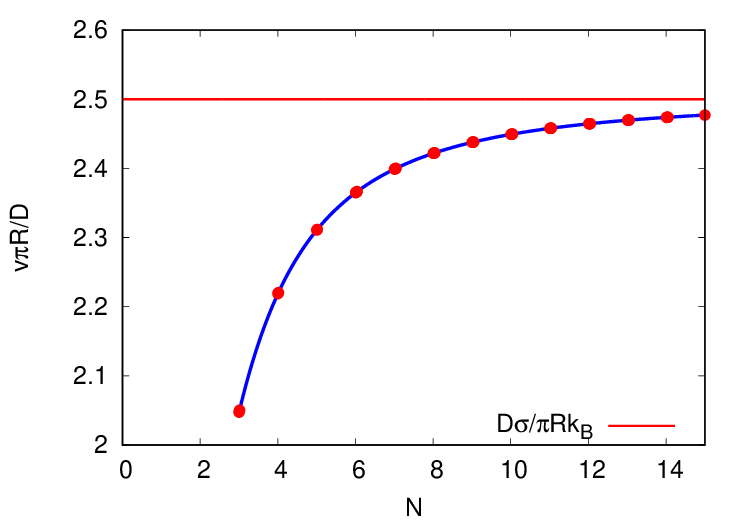}
\caption{\label{fig5} Velocity $v$ plotted against the number $N$ of sites per cycle by keeping the entropy $\sigma$ per cycle and the diffusion coefficient $D$ constant for $N\geq 3$. We choose the produced entropy per cycle to be $\sigma/k_B=5$. The red line is the value $v(N\to \infty)\to D\sigma /\pi R k_B$ in the limit of $N\to \infty$.}
\end{figure}

Next, consider case (B). Eq.\ \eqref{eq9} with Eqs.\ \eqref{b1} and \eqref{b2} leads to the velocity on $N$
\begin{align}
\label{velocityA}
v=(\alpha-\beta)\frac{2\pi R}{N}=\frac{DN}{\pi R}\tanh{\bigg[ \frac{\sigma}{2Nk_B}\bigg]}.
\end{align}
The velocity $v$ increases with $N$ and so the time of a full completion of the system cycle $\tau=2\pi R/v$ is reduced and minimizes for $N\to\infty$, see Fig.\ \ref{fig5}.

\section{Randomness parameter and variance in cycle completion time}
\label{randomness}
In the previous Sections \ref{SecIII} and \ \ref{IV}, we have analyzed, for a biased random process, the dependence of different physical observables on the number intermediate sites $N$ taken to complete a given cycle of operation, as a way to demonstrate the non-uniqueness of going to the continuous limit ($N\to \infty$) of this process. Another perspective of this problem has been studied in Refs. \cite{svo1994,sch1995,tho2001,BookStat}, where the values of physical observables associated with enzyme catalyzed cycles were used to set bounds on the number of intermediate cycle steps \cite{footnote6}. It was pointed out that in addition to the average speed $v$ and diffusion coefficient $D$, their ratio provides important information on the observed walk statistics \cite{svo1994,sch1995,tho2001,BookStat}.
Explicitly, we characterize the random process in terms of their forward and backward rates, $\alpha$ and $\beta$, respectively, and an equidistant step length $\Delta x$. If the process starts at $n=1$ and a random site $n$ is reached at time $t$, then the process will be on average at site $\langle n \rangle =vt/\Delta x$, whilst the random diffusive process produces a variance in the site by $\langle \delta n^2 \rangle \equiv \langle n^2 \rangle-\langle n \rangle^2=2Dt/\Delta x^2$, see Eqs.\ \eqref{eq9} and \eqref{eq9b} in the limit $t\to \infty$. 
These two quantities can be combined into a randomness parameter, which for the given step size $\Delta x$ reads \cite{svo1994,sch1995,tho2001,BookStat}
\begin{align}
\label{random}
r\equiv\frac{\langle \delta n^2 \rangle}{\langle n \rangle}=\frac{2D}{\Delta x v}=\frac{\alpha+\beta}{\alpha-\beta},
\end{align}
where $v$ and $D$ are defined by Eqs.\ \eqref{eq9} and \eqref{eq9b}, respectively.
Alternatively, we may consider the random passage time $\tau$ at which, starting from $n=1$, the walk reached for the first time the site $N$, namely a distance $N\Delta x$ from the starting point. For walks of uniform step length and finite bias, it has been shown \cite{sch1995,svo1994} that for large enough $N$, the randomness parameter can be expressed in terms of the first two moments of the passage time distribution
\begin{align}
\label{random3}
r= \frac{\langle \tau ^2\rangle - \langle \tau \rangle^2}{\langle \tau \rangle^2}=\frac{\langle \delta \tau^2 \rangle}{\langle \tau \rangle^2},
\end{align}
where $\langle \tau \rangle$ is the average time for a cycle completion and $\langle \delta \tau^2 \rangle$ its variance. Note that for many enzyme reaction cycle, the backward reaction rates are often sufficiently low as to be negligible. In such cases the pathway consists of a sequence of $N$ forward reactions only and the randomness parameter ($r=N_{min}^{-1}$) can be used to estimate the minimal number of kinetic sites that compose the underlying kinetic model \cite{sch1995,BookStat}.
In general, when considering forward and backward steps and using the average cycle completion time $\langle \tau \rangle$, we can calculate the variance in cycle completion time to
\begin{align}
\label{var}
\langle \delta \tau ^2 \rangle= r \langle \tau \rangle^2.
\end{align}

We can now apply the results of Secs.\ \ref{SecIII} and \ref{IV} to examine the behavior of these observables in our different limiting cases. In our case $N$ corresponds to the number of sites per cycle and, consequently, $\tau$ is the time for the process to complete the cycle. Increasing $N$ is done with eventually approaching a continuous description, so the cycle length $N\Delta x=2\pi R$ is kept fixed.

Consider first the condition of a constant velocity and diffusion constant by increasing the number of sites $N$, see Sec.\ \ref{SecIII}. Using Eqs.\ \eqref{eq12} and \eqref{eq13} in \eqref{random}, the randomness parameter is
\begin{align}
\label{random1}
r=\frac{D}{\pi R v} N
\end{align}
and is linear in $N$.
\\ Next, for the condition of constant entropy production $\sigma$ per cycle under either constant velocity $v$ or constant diffusion $D$, see Sec.\ \ref{IV}, we find in both cases, using the respective rates in Eqs.\ \eqref{a1} and \eqref{a2} or Eqs.\ \eqref{b1} and \eqref{b2}, the randomness parameter to be
\begin{align}
\label{random2}
r=\coth{\bigg[\frac{\sigma}{2Nk_B}\bigg]}.
\end{align}
The randomness parameter \eqref{random2} is determined by the number of sites $N$ and the thermodynamic entropy production or heat dissipation into the environment which equals, when neglecting the movement against an external force, the energy drop per cycle, see Sec.\ \ref{unicylcic}. Note that $r$ tends to infinity in the limit $\sigma\to 0$ and in the continuous limit $N\to \infty$ since both limits reflect the equilibrium situation where forward and backward rates will be alike. In the limit $\sigma\to \infty$, given finite $N$, we find $r=1$, which is expected for the so-called "Poisson" motion since the infinite energy drop per cycle leads to an unidirectional motion \cite{svo1994}.
\begin{figure}[h!!!!!]
\centering
\includegraphics[width=\linewidth]{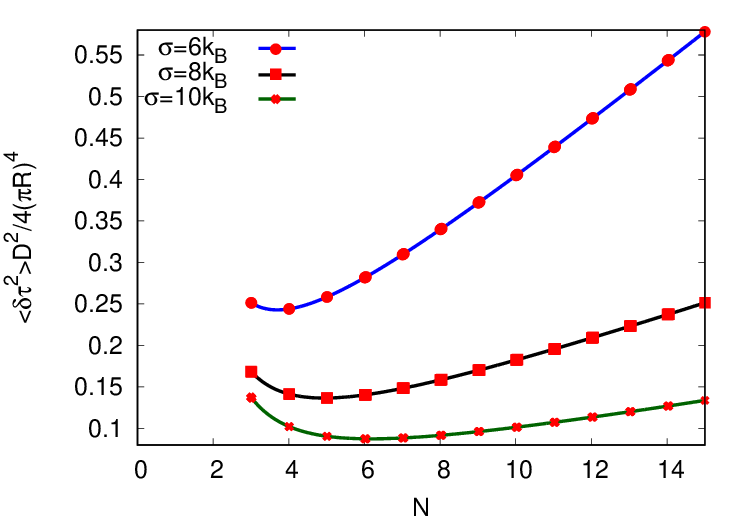}
\caption{\label{fig5b} Variance in cycle completion time $\langle \delta \tau ^2 \rangle$ plotted against the number $N$ of sites per cycle by keeping the entropy $\sigma$ per cycle and the diffusion coefficient $D$ constant for $N\geq 3$ while the velocity $v(N)$ results from Eq.\ \eqref{velocityA}. We show $\langle \delta \tau ^2 \rangle$ for three different choices of $\sigma$.}
\end{figure}

With the randomness parameter at hand we can now study the variance in the cycle completion time (Eq.\ \eqref{var}). It has been shown that for a biased random walk for large $N$ that the average completion time $\langle \tau \rangle=N(\alpha-\beta)^{-1}$ \cite{kha1983}. Consider first the cases (A) of keeping the velocity and diffusion coefficient constant as well as (B) keeping the entropy production per cycle and velocity constant. We find for (A) given Eq.\ \eqref{random1}, together with Eqs.\ \eqref{eq12} and \eqref{eq13} in Eq.\ \eqref{var} where $\langle \tau \rangle=N(\alpha-\beta)^{-1}$ the variance in cycle completion time to be
\begin{align}
\label{var1}
\langle \delta \tau^2 \rangle = \frac{4D \pi R}{v^3} N,
\end{align}
and, equivalent for case (B) by using Eq.\ \eqref{random2}, together with Eqs.\ \eqref{a1} and \eqref{a2} in Eq.\ \eqref{var} 
\begin{align}
\label{var2}
\langle \delta \tau^2 \rangle = \coth{\bigg[\frac{\sigma}{2Nk_B}\bigg]}\frac{4\pi^2R^2}{v^2}.
\end{align}
In both Eqs.\ \eqref{var1} and \eqref{var2}  the variance in cycle completion time increases monotonously with site number $N$. This reflects the fact that with increasing number of sites $N$ per cycle the inter-site rates become more alike which increases the overall "randomness", and, thus, $\langle \delta \tau ^2 \rangle$ for the total cycle completion. 

In contrast when $N$ is changed while keeping a constant diffusion coefficient and entropy production, Eq.\ \eqref{random3}, together with \eqref{velocityA} and \eqref{random2} lead to
\begin{align}
\langle \delta \tau ^2 \rangle = \coth^3{\bigg[\frac{\sigma}{2Nk_B}\bigg]}\frac{4(\pi R)^4}{D^2N^2}.
\end{align}
Interestingly, $\langle \delta \tau ^2 \rangle$ goes through a minimum with increasing site number $N$, see Fig.\ \ref{fig5b}. It should be kept in mind however that Eq.\ \eqref{random3} and, consequently, Eq.\ \eqref{var}, was derived under the assumption $N$ is large so that this observation should not be regarded conclusive.

\section{N-site process with time step discretization}
\label{V}

Consider now the same biased random walk process in which a cycle of length $2\pi R$ is traversed in $N$ steps, so that $N\Delta x=2\pi R$, but where the system is restricted to move (by intersite distance $\Delta x$) only at finite time intervals $\Delta t$. Indeed, small systems which are periodically driven can be thought of as discrete-time processes, see \cite{ros2017}. As shown below, $\Delta x$ and $\Delta t$, are not independent of each other but some freedom exists in their choices. $N\Delta E$ is the energy drop per such cycle (see Sec.\ \ref{unicylcic}, recall that $\Delta E$ determines the detailed balance ratio of the forward and backward rates according to Eq.\ \eqref{eqLocal}). The probability to be at site $n$, namely at position $x=n\Delta x$ on the cycle at time $t=M\Delta t$, is governed by the Makrov chain \cite{footnote4,KlafterBook}
\begin{align}
\label{eq1}
 P(x,t+\Delta t)&= \alpha \Delta t P(x-\Delta x,t)+\beta \Delta t P(x+\Delta x,t)\\ \notag &+(1-\alpha \Delta t -\beta\Delta t)P(x,t).
\end{align}

Here, $\alpha \Delta t$ and $\beta \Delta t$ are the probabilities (both assumed linear in $\Delta t$) to move a step forward and backward, respectively. Note that $(\alpha+\beta)\Delta t\leq 1$ has to be imposed in Eq.\ \eqref{eq1} to ensure positivity. As before, the process has periodic bounderies so that, after the final site $n=N$ has reached, it restarts at the beginning $n=1$ and its original energy is restored by some external work reservoir between sites $N$ and $N+1=1$ (see discussion in Sec.\ \ref{unicylcic}). We use this model to study the effect of time discretization on dynamical properties of the process. To calculate the velocity $v$ and diffusion coefficient $D$, we determine the generating function $P(s,t)=\sum_x s^x P(x,t)$ where the moments can be calculated by $\langle x^m \rangle = (s \partial/\partial s)^m P(s,t)|_{s=1}$. 
For the initial condition $P(x=0,t=0)=1$ (so that $P(s,t=0)=1$) we find the generating function to be
\begin{align}
P(s,t=&M\Delta t) \\ \notag &= [\alpha \Delta t s^{\Delta x} + \beta \Delta t s^{-\Delta x}+(1-\alpha\Delta t -\beta \Delta t)]^M.
\end{align}

The velocity and diffusion coefficient are determined as follows. Starting at $x=0$, we find in the long time limit (at steady state) $t\to \infty$ (see details in Appendix \ref{disc})
\begin{align}
\label{velocity3}
v &= \frac{\langle x \rangle }{t}= (\alpha- \beta) \Delta x ; \\
\label{Diffusion2}
2D&= \frac{\langle x^2 \rangle -\langle x \rangle^2}{t}= (\alpha + \beta) \Delta x^2 - (\alpha \Delta x - \beta \Delta x)^2 \Delta t \\ \notag &=(\alpha + \beta) \Delta x^2 - v^2 \Delta t.
\end{align}

The velocity $v$, Eq.\ \eqref{velocity3}, is the same as in the continuous time case (Eq.\ \eqref{eq9}), whereas the diffusion coefficient in Eq.\ \eqref{Diffusion2} is smaller by $v^2\Delta t$ in comparison to the continuous time case of Eq.\ \eqref{eq9b}. Refs. \cite{koz1999,chi2018} associate the bigger variance in the continuous-time master equation with higher fluctuations in the total number of hops observed in a given time interval. 
Next, consider the process as $N$ increases. As in Secs.\ \ref{SecIII} and \ref{IV}, we may consider an increase of $N$ while keeping $v$ and $D$ constant or while keeping only one of them together with $N\Delta E$ constant. As examples of the effect of moving in discrete time steps, we study the cases (A) constant $v$ and $D$ and (B) constant $v$ and $N\Delta E$.

(A) Keeping $v$ and $D$ constant, we scale the rates again analog to Eqs.\ \eqref{a1} and \eqref{a2} with the total site number $N$ (given by the intersite distance $\Delta x=2\pi R/N$)
\begin{align}
\label{a11}
\alpha &=\frac{(D+v^2\Delta t/2)N^2}{(2\pi R)^2} + \frac{v N}{4\pi R},
\\
\label{a22}
\beta &=\frac{(D+v^2\Delta t/2)N^2}{(2\pi R)^2} - \frac{v N}{4\pi R}.
\end{align}
Note that the modification of the rates $\alpha$ and $\beta$ depends on $\Delta t$. As before, this rescaling implies also a change in $\Delta E$ (see discussion in Sec.\ \ref{SecIII}) so that the detailed balance relation is maintained. The condition $(\alpha+\beta)\Delta t\leq 1$ in Eq.\ \eqref{eq1} together with Eqs.\ \eqref{a11} and \eqref{a22} restricts the choices for $\Delta t$ given $\Delta x=2\pi R/N$ to 
\begin{align}
\label{deltat}
0\leq \Delta t \leq \frac{-D}{v^2}+\sqrt{\frac{D^2}{v^4}+\frac{\Delta x^2}{v^2}},
\end{align}
which implies that $\Delta t$ and $\Delta x$ cannot be assigned independently of each other. In the limit $N\to \infty$ ($\Delta x \to 0$), this inequality (Eq.\ \eqref{deltat}) becomes $0\leq \Delta t \leq \Delta x^2/2D$.

\begin{figure}[h!!!!!]
\centering
\includegraphics[width=\linewidth]{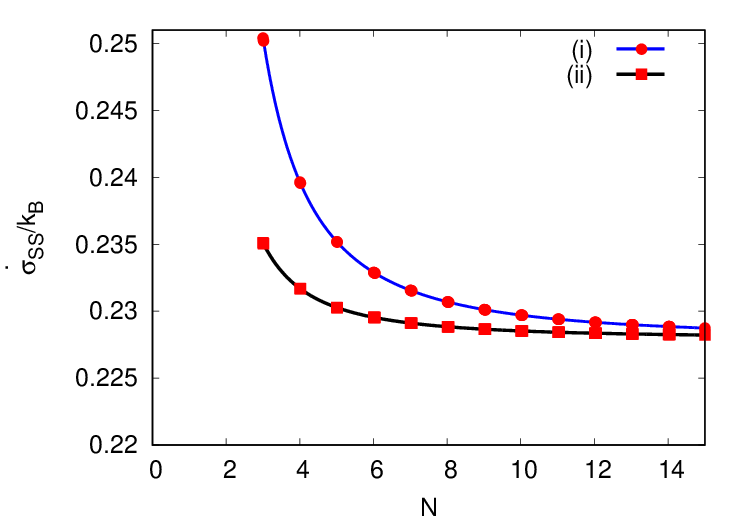}
\caption{\label{fig4} Entropy production rate at steady state, $\dot{\sigma}_{SS}$, against the number per cycle $N$ by keeping the velocity $v$ and diffusion coefficient $D$ constant. Here, $v$ (chosen to be $v=\frac{3D}{2\pi R}$) needs to be within the bound implied by $\beta\geq 0 $ of Eq.\ \eqref{a22} for $N\geq 3$. (i) The blue circled line represents the entropy production rate in the continuous time limit $\Delta t\to 0$. (ii) The black squared line is the entropy production rate for a discrete time process using the maximal $\Delta t$ allowed by Eq.\ \eqref{deltat}, $\Delta t =-\frac{D}{v^2}+\sqrt{\frac{D^2}{v^2}+\frac{\Delta x^2}{v^2}}$.}
\end{figure}

Consider next the entropy production for this discrete hopping process. The average entropy change per step is
\begin{align}
\label{entropychange}
\Delta \sigma = k_B \alpha \ln \bigg [ \frac{\alpha}{\beta} \bigg ] + k_B \beta \ln \bigg [ \frac{\beta}{\alpha} \bigg ],
\end{align}
where the two terms represent the entropy change in a forward and backward step multiplied by the probabilities that the respective step occurs. The rate of entropy change at steady state (entropy production rate) is given by
\begin{align}
\label{entropyincrement}
\dot{\sigma}_{SS}&=k_B\frac{\Delta \sigma}{\Delta t}=k_B (\alpha - \beta) \ln{\bigg [ \frac{\alpha}{\beta}\bigg ]}
\\ \notag &=\frac{k_B N}{2 \pi R}  v  \ln{\bigg [ \frac{\frac{(D+v^2\Delta t/2) N^2}{(2 \pi R)^2} + v\frac{N}{4\pi R}}{\frac{(D+v^2\Delta t/2) N^2}{(2 \pi R)^2} - v\frac{N}{4\pi R}}\bigg ]}.
\end{align}
Interestingly, comparing the resulting expression \eqref{entropyincrement} to its analog \eqref{heatsteadystate} for the continuous master equation, we obtain a similar result, but with an additional term $v^2\Delta t/2$ added to the diffusion constant. The additional term effectively modifies the TUR relation (Eq.\ \eqref{TUR}) as the relative uncertainty $2D/v^2$ changes. A similar observation was made in Ref.\ \cite{pro2017}. The resulting entropy production rate in Eq.\ \eqref{entropyincrement} for given site number $N$ per cycle is reduced if we choose a finite $\Delta t$ (given the restriction on choices of $\Delta t$ by Eq.\ \eqref{deltat}), see Fig.\ \ref{fig4}. This might be understood since during a given time interval the variance in position $x$ on the cycle is reduced (Eq.\ \eqref{Diffusion2}) by allowing intersite hops only in intervals $\Delta t$. 
In the continuous limit ($N \to \infty$, $\Delta x \to 0$ and $\Delta t\to 0$), however, Eq.\ \eqref{entropyincrement} yields
\begin{align}
\dot{\sigma}_{SS} = k_B  \bigg[\frac{ v^2}{D}\bigg ],
\end{align}
which is the same result obtained in this limit in Sec.\ \ref{SecIII} (see Eq.\ \eqref{heatsteadystate}).

(B) Consider next the dependence on $N$ under the condition of a constant energy drop $N\Delta E$ per cycle and constant velocity $v$. This is the analogue consideration as in Sec.\ \ref{IV}, but where the intersite hops are only allowed at time intervals $\Delta t$ (that are restricted by the given $N$ according to Eq.\ \eqref{deltat2} below). We assume that all energy invested into the system is dissipated as heat into the environment, so the steady state entropy production per cycle is
\begin{align}
\sigma=N\frac{\Delta E}{T}=k_B N \ln \bigg ( \frac{\alpha}{\beta}\bigg),
\end{align}
where $N$ is the total number per cycle. 

Given the constant velocity, Eq.\ \eqref{velocity3}, the forward and backward rates are equivalent to Eqs.\ \eqref{a1} and \eqref{a2}

\begin{align}
\label{a1b}
\alpha &= \frac{vN}{2\pi R\big(1-e^{\frac{-\sigma}{Nk_B}}\big)}, \\
\label{a2b}
\beta &= \frac{vN}{2\pi R\big(e^{\frac{\sigma}{Nk_B}}-1\big)}.
\end{align}
The restriction $(\alpha+\beta)\Delta t\leq 1$ in Eq.\ \eqref{eq1} together with Eqs.\ \eqref{a1b} and \eqref{a2b}, limits the choices for $\Delta t$ given $\Delta x=2\pi R/N$ to
\begin{align}
\label{deltat2}
0\leq \Delta t \leq \frac{\Delta x}{v} \tanh\bigg[\frac{\sigma}{2Nk_B}\bigg].
\end{align}
In the limit $N\to \infty$ ($\Delta x \to 0$ and $\Delta t \to 0$), the inequality in Eq.\ \eqref{deltat2} becomes $0\leq \Delta t \leq \Delta x \sigma /(vNk_B)$.

\begin{figure}[h!!!!!]
\centering
\includegraphics[width=\linewidth]{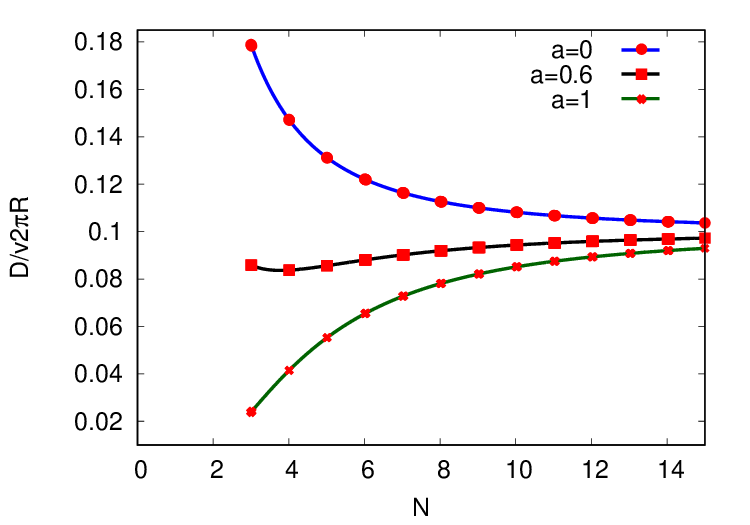}
\caption{\label{fig4b} The diffusion coefficient $D$ plotted against the number of sites $N$ per cycle by keeping the entropy production per cycle $\sigma$ and velocity $v$ constant. The chosen time interval $\Delta t$ for given $N$ is restricted by the inequality \eqref{deltat2}. We take $\Delta t(a,N)=a \frac{\Delta x}{v}\tanh\big[ \frac{\sigma}{2Nk_B} \big]$ (where $0\leq a \leq 1$) and depict $D(a,N)$ with $a=0$ (blue curve), $a=0.6$ (black curve) and $a=1$ (green curve). We choose the produced entropy per cycle to be $\sigma/k_B=10$ for $N\geq 3$.}
\end{figure}

For a given $N$ and $\Delta x=2\pi R/N$, the time step $\Delta t$ needs to satisfy the inequality \eqref{deltat2}.  Here we take 
\begin{align}
\label{deltat3}
\Delta t = a \frac{\Delta x}{v}\tanh\big[ \frac{\sigma}{2Nk_B}\big]
\end{align}
with $0\leq a \leq 1$ and use Eq.\ \eqref{Diffusion2} to get
\begin{align}
\label{diffusion3}
D&=\frac{1}{2}\bigg( (\alpha+\beta)\Delta x^2-v^2\Delta t\bigg)\\ \notag &=\frac{v\pi R}{N}\bigg ( \coth\bigg[ \frac{\sigma}{2Nk_B}\bigg]-a\tanh\bigg[ \frac{\sigma}{2Nk_B}\bigg] \bigg).
\end{align}
Interestingly, the dependence on time discretization translates here to a dependence of $D$ on the choice of $a$, see Fig.\ \ref{fig4b}. Given $N$ for different finite time $\Delta t$ (scaled between $0\leq a \leq 1$) the diffusion coefficient, and consequently, the variance in $x=n\Delta x$, are strongly reduced for increasing $\Delta t$ (less fluctuations in the total number of transitions for finite $\Delta t$). As expected, in the limit $N\to \infty$ and $\Delta t\to 0$, the diffusion constant takes the form $D=v/Fk_BT=\mu k_B T$ with the mobility $\mu=v/F$ and the corresponding force $F=T\sigma/2\pi R$, \cite{ein1905,smo1906}, see discussion in Sec.\ \ref{IV}. Therefore, given the chosen $\Delta t$ and its above discussed effect on $D$, the latter increases or decreases with $N$ to the final value $D=\mu k_B T$ as depicted in Fig.\ \ref{fig4b}.

To summarize this section, when describing the dynamics of a (cyclic) process in discrete time intervals, thermo-dynamical properties of the process, e.g., the entropy production or diffusion coefficient, are affected by this time discretization. The discretization in time, however, cannot be chosen arbitrarily but must obey the bounds given by the state-space discretization of the process. If the process dynamics is going to a continuous state-space dynamics ($N\to \infty$), the time evolution needs to be described by intervals $\Delta t\to 0$, i.e., equivalently by a continuous time scale and all effects vanishes.

\section{Conclusion}
\label{conc}
In this paper we have investigated a cyclic and uni-thermal thermodynamic process by a model of a biased random walk between $N$ sites on a cycle of a given length by an exact solvable master equation. We note that many dynamical site (state)-space processes with periodic boundary conditions may be equivalently mapped to this process. The limit $N\to \infty$ corresponds to the continuous limit that is usually captured by a Fokker-Planck equation. This limit is taken by keeping low order moments of system observables, i.e., the velocity $v$ and diffusion coefficient $D$ constant. We have shown that the entropy produced, or, equivalently, the energy drop per cycle is reduced when going to this continuous description. This has direct consequences for the efficiency of transferring input power into useful output power when an opposing force acts on the cyclic process. In particular, more power can be extracted from the process with increasing number of sites $N$ per cycle length.

An important outcome of our analysis is that the procedure in going to the continuous description of the process is not unique and depends on the physical observables that are assumed to be invariant under this limiting process. In addition to taking the limit $N\to \infty$ while keeping $v$ and $D$ constant we have also analyzed this limiting process while keeping $v$ or $D$ and $N \Delta E$ constant. Interestingly, considering the limiting process under constant $v$ and $N \Delta E$, we have shown that diffusion coefficient $D$ for finite cycle velocity $v$ in the limit $N \to \infty $ takes the same value as in linear response $v \to 0$ limit. Additionally, when analyzing the cycle randomness statistics and, in particular, the variance in the cycle completion time $\langle \delta \tau^2 \rangle$. We find that with increasing $N$, $\langle \delta \tau^2 \rangle$ increases signaling the increasing randomness in the cycle. Finally, we have studied the dependence on $N$ in the case where the transitions between sites are only allowed at fixed time intervals $\Delta t$. We find that not only the entropy production rate per cycle (when $v$ and $D$ are kept constant), but also the diffusion coefficient $D$ (when $v$ and $N\Delta E$ are kept constant) are strongly affected by the way time discretization is introduced for a given $N$. 

In conclusion, on can use the total site number $N$ as a control parameter to design "usefull" physical and thermodynamic (cycle) processes by keeping desirable observables constant and affecting others. It may provide valuable insights into the engineering of small (molecular) machines capable of performing specific tasks with high efficiency and precision. Further investigations of the discrete to continuous transition in state-space and its potential impact on information-to-work conversion are kept future research.

\begin{acknowledgments}
This work has been supported by the U.S. National Science Foundation under the Grant No. CHE1953701 and the University of Pennsylvania.
\end{acknowledgments}

\bibliography{MS}

\appendix

\section{Three-site cycle}
\label{three}

Consider a three site system with sites $1,2$ and $3$ where the transition rates $k_{1\to 2}=k_{2\to 3}=k_{3\to 1}=\alpha$ and $k_{2\to 1}=k_{1\to 3}=k_{3\to 2}=\beta$.
The master equation can be written as
\begin{align}
\begin{pmatrix}
\dot{P}_1\\
\dot{P}_2 \\
\dot{P}_3
\end{pmatrix}
=
\begin{bmatrix}
-(\alpha+\beta) & \beta & \alpha\\
\alpha & -(\alpha+\beta) & \beta \\
\beta & \alpha & -(\alpha+\beta) \\
\end{bmatrix}
\begin{pmatrix}
P_1\\
P_2 \\
P_3
\end{pmatrix}.
\end{align}
Assuming the initial condition $P(t=0)=(1,0,0)$, the probabilities to be on site $n$ have the solution
\begin{align}
P_1(t)&=\frac{1}{3}\bigg( 1+ 2 e^{-3\Phi t} \cos(\sqrt{3}\Psi t) \bigg),\\
P_2(t)&=\frac{1}{3}\bigg( 1 - 2 e^{-3\Phi t} \cos(\sqrt{3}\Psi t-\pi/3) \bigg),
\\
P_3(t)&=\frac{1}{3}\bigg( 1 - 2 e^{-3\Phi t} \cos(\sqrt{3}\Psi t+\pi/3) \bigg),
\end{align}
where $\Phi=(\alpha+\beta)/2$ and $\Psi=(\alpha-\beta)/2$.

The entropy production rate for the three site system reads
\begin{align}
\dot{\sigma}(t)&=(\alpha-\beta)\ln{\frac{\alpha}{\beta}}+(P_1(t)\alpha-P_2(t)\beta)\ln\bigg(\frac{P_1(t)}{P_2(t)}\bigg)\\ \notag &+(P_2(t)\alpha-P_3(t)\beta)\ln\bigg(\frac{P_2(t)}{P_3(t)}\bigg)\\ \notag &+(P_3(t)\alpha-P_1(t)\beta)\ln\bigg(\frac{P_3(t)}{P_1(t)}\bigg),
\end{align}
and the (constant) entropy flow
\begin{align}
\dot{S}_e(t)\equiv\dot{S}_e &=-(\alpha-\beta)\ln{\frac{\alpha}{\beta}}.
\end{align}
At stationary state, the entropy production and flow satisfy $\dot{\sigma}=-\dot{S}_e$.

\section{Four-site cycle}
\label{four}

With the same methodology as before but now with 4 sites in a cycle we find
\begin{align}
P_1(t)&=\frac{1}{4}\bigg( 1+e^{-2\Phi t}+ 2 e^{-\Phi t} \cos(\Psi t) \bigg),\\
P_2(t)&=\frac{1}{4}\bigg( 1 - e^{-2\Phi t}+ 2 e^{-\Phi t} \sin(\Psi t) \bigg),
\\
P_3(t)&=\frac{1}{4}\bigg( 1 +e^{-2 \Phi t}- 2 e^{-\Phi t} \cos(\Psi t) \bigg),
\\
P_3(t)&=\frac{1}{4}\bigg( 1 -e^{-2 \Phi t}- 2 e^{-\Phi t} \sin(\Psi t) \bigg),
\end{align}
where $\Phi=(\alpha+\beta)$ and $\Psi=(\alpha-\beta)$.
The entropy production rate for the four site system reads
\begin{align}
\dot{\sigma}(t)&=(\alpha-\beta)\ln{\frac{\alpha}{\beta}}+(P_1(t)\alpha-P_2(t)\beta)\ln\bigg(\frac{P_1(t)}{P_2(t)}\bigg)\\ \notag &+(P_2(t)\alpha-P_3(t)\beta)\ln\bigg(\frac{P_2(t)}{P_3(t)}\bigg) \\ \notag &+(P_3(t)\alpha-P_4(t)\beta)\ln\bigg(\frac{P_3(t)}{P_4(t)}\bigg)\\ 
\notag &+(P_4(t)\alpha-P_1(t)\beta)\ln\bigg(\frac{P_4(t)}{P_1(t)}\bigg),
\end{align}
and the (constant) entropy flow
\begin{align}
\dot{S}_e&=-(\alpha-\beta)\ln{\frac{\alpha}{\beta}}.
\end{align}
At stationary state, the entropy production and flow satisfy $\dot{\sigma}=-\dot{S}_e$.

\section{N-site cycle}
\label{N-site cycle}

We perform a discrete Fourier transform of the master equation \eqref{master1} we obtain with $P(s,t)= \sum_{n=1}^N z_s^n P(n,t)$ where $z_s=\exp{(2\pi i s/N)}$. We get
\begin{align}
\dot{P}(s,t)&=[\alpha e^{2\pi i s/N}+ \beta e^{-2\pi i s/N}-(\alpha+\beta)]P(s,t) \\ \notag
&=[(\alpha+\beta) \cos(2\pi i s/N) + i(\alpha-\beta) \sin(2\pi i s/N) \\ \notag &-(\alpha+\beta)]P(s,t).
\end{align}
Its solution reads
\begin{align}
\label{eq36}
P(s,t)&=e^{[(\alpha+\beta) \cos(2\pi i s/N) + i(\alpha-\beta) \sin(2\pi i s/N)-(\alpha+\beta)]t} \\ \notag & \times P(s,t=0),
\end{align}
where we have assumed $P(s,t=0)=\sum_{n=1}^N z_s^n P(n,t=0)$.
We now perform a little algebraic transformation in Eq.\ \eqref{eq36}
\begin{align}
P(s,t)&=e^{\sqrt{\alpha \beta}t\big((\sqrt{\frac{\alpha}{\beta}}+\sqrt{\frac{\beta}{\alpha}}) \cos(2\pi i s/N) + i(\sqrt{\frac{\alpha}{\beta}}-\sqrt{\frac{\beta}{\alpha}}) \sin(2\pi i s/N)\big)} \\ \notag & \times e^{-(\alpha+\beta)t}P(s,t=0) \\ \notag
&=e^{\sqrt{\alpha \beta}t\big(\sqrt{\frac{\alpha}{\beta}}e^{2\pi i s/N}+\sqrt{\frac{\beta}{\alpha}} e^{-2\pi i s/N} \big)} e^{-(\alpha+\beta)t}P(s,t=0) \\ \notag
&=e^{\big[\alpha (z_s-1)+\beta \big(\frac{1}{z_s}-1\big)\big]t}P(s,t=0)=e^{\lambda_s t}P(s,t=0).
\end{align}
The Fourier back transform then reads
\begin{align}
\label{back}
P(n,t)&=\frac{1}{N} \sum_{s=1}^N z_s^{-n}e^{\lambda_s t}P(s,t=0),
\end{align}
where $z_s=\exp{(2\pi i s/N)}$. 
The stationary state is reached for $t\to \infty$. In this case only the term with $s=N$ in Eq.\ \eqref{back}. The mode $s=1$ corresponds to $z_N=1$ and $\lambda_s=0$. In this limit 
\begin{align}
\label{back}
P(n,t \to \infty)=\frac{1}{N} P(N,t=0)=\frac{1}{N} \sum_{n=1}^N P(n,t=0)=\frac{1}{N},
\end{align}
where we assume to start a specific site, say $n=1$ at $t=0$, so $P(n=1,t=0)=1$. The probabilities in the stationary state become constant and independent of the site $n$. The other eigenvalues $\lambda_s$ determine the relaxation time of the mode $s$ and, thus, how quick the stationary state is reached.
For $N\gg 1$ and finite $s$ we find
\begin{align}
\lambda_s &=\alpha (z_s-1)+\beta \big(\frac{1}{z_s}-1\big) \\ \notag &\simeq (\alpha -\beta) \frac{2\pi i s}{N}-(\alpha + \beta) \frac{4\pi^2 s^2}{N^2}.
\end{align}

\section{Discretization in time space}
\label{disc}
To calculate the velocity $v$ and diffusion constant $D$, we determine the generating function $P(s,t)=\sum_x s^x P(x,t)$ where the moments can be by $\langle x^n \rangle = (s \partial/\partial s)^n P(s,t)|_{s=1}$. 
For the initial condition $P(s,t=0)=1$ we find the generating function to be
\begin{align}
P(s,t)= [\alpha \Delta t s^{\Delta x} + \beta \Delta t s^{-\Delta x}+(1-\alpha\Delta t -\beta \Delta t)]^M,
\end{align}
where $t=M\Delta t$. 

The first moment can be determined to be
\begin{align}
\langle x \rangle = (\alpha - \beta) \Delta x \Delta t M = (\alpha - \beta) \Delta x t.
\end{align}
The second moment reads
\begin{align}
\langle x^2 \rangle &= (\alpha \Delta x^2 + \beta \Delta x^2) \Delta t M \\ \notag &+ (\alpha \Delta x - \beta \Delta x)^2 \Delta t^2 M(M-1) \\   \notag 
&= (\alpha + \beta) \Delta x^2 t + (\alpha \Delta x - \beta \Delta x)^2 (t^2-t\Delta t).
\end{align}
The variance can then be found to be
\begin{align}
\langle x^2 \rangle -\langle x \rangle^2 = (\alpha +\beta) \Delta x^2 t - (\alpha \Delta x - \beta \Delta x)^2 t\Delta t.
\end{align}

\clearpage


\end{document}